%% file: main.tex
\documentclass[runningheads]{llncs}
\usepackage[T1]{fontenc}
\usepackage{graphicx}
\usepackage{booktabs}

\begin{document}
\title{CLARA: An AI-Augmented Analytics Dashboard for Collaboration Literacy\thanks{Accepted at AIED 2026; to appear in Springer LNAI/LNCS.}}
\author{Dawei Xie\inst{1} \and
Khalil Anderson\inst{1} \and
Tochukwu Eze\inst{1} \and
Chenghong Lin\inst{1}\and
Bookyung Shin\inst{1} \and
Marcelo Worsley\inst{1}}

\authorrunning{D. Xie et al.}
\institute{Northwestern University, Evanston, IL 60208, USA \\
\email{\{daweixie, khanders, td.eze, chenghonglin, bookyungshin\}@u.northwestern.edu} \\
\email{marcelo.worsley@northwestern.edu}}
\maketitle              % typeset the header of the contribution
\begin{abstract}
Collaboration literacy requires adapting to the evolving demands of group work within complex discussions, making it difficult to develop and assess. Traditional analytics metrics capture behavioral signals while missing the semantic dimensions of how learners approach collaboration and build on each other's ideas. We present Collaboration Literacy through Artifact Reasoning and Augmentation (CLARA), an agentic analytics system that extracts semantic representations from transcripts as analytics artifacts: concept maps representing emergent ideas and relationships, and collaboration assessment characterizing collaboration quality across seven dimensions. While users explore these artifacts through the dashboard, the same artifacts are indexed into distinct vector database collections for agent retrieval and reasoning. This architecture establishes a human-AI common ground where users and AI can operate over shared representations. Evaluation results show that CLARA produces reliable collaboration quality analysis and, owing to the artifacts serving as knowledge infrastructure, improves both retrieval performance and response quality over transcript-only baselines. Our work suggests that AI-produced artifacts may scaffold human interpretation and ground AI reasoning in learning analytics workflows.
\end{abstract}
\input{1_introduction}

\input{2_related_work}
\input{3_system_design}
\input{4_evaluation}
\input{5_discussion}
\input{6_conclusion}

\begin{credits}
%\subsubsection{\ackname} We would like to thank all anonymous reviewers %for their dedicated reviews and comments.

\subsubsection{\discintname}
The authors have no competing interests to declare that are relevant to the content of this article.
\end{credits}
\bibliographystyle{splncs04}
\bibliography{aied2026}

\end{document}

%% file: 1_introduction.tex
\section{Introduction}
Collaboration literacy, the capacity to adapt to the evolving demands of group work~\cite{worsley2020towards}, remains a persistent challenge in learning analytics. Collaborative group work often involves in-depth discussions in which concepts evolve and are negotiated~\cite{stahl2004building}, and the complexity of these interactions makes collaboration quality difficult to assess through behavioral signals alone. Understanding how learners coordinate to approach collaboration and build on each other's ideas requires capturing semantic dimensions of discourse that traditional analytics approaches may fall short of representing.

Large language models (LLMs) offer a path forward as they are capable of extracting structured representations from conversations. While many applications treat LLMs as endpoints that summarize or answer questions, we propose a different framing: LLMs as producers of multiple semantic artifacts that become infrastructure for downstream retrieval, reasoning, and human-AI sensemaking. This paper presents \textbf{C}ollaboration \textbf{L}iteracy through \textbf{A}rtifact \textbf{R}easoning and \textbf{A}ugmentation (CLARA), an agentic analytics system for collaborative literacy. The system extracts multiple analytics artifacts from discussion transcripts: concept maps in which nodes represent emergent ideas and edges represent conceptual relationships; 7C collaboration assessments that characterize discussions along seven dimensions; and psycholinguistic metrics that track textual sentiment over time. All artifacts, including raw transcripts, are embedded and indexed into distinct database collections. An AI agent then operates over these heterogeneous representations, surfacing patterns across sessions, and synthesizing grounded feedback in response to user queries.

Our work makes three contributions: (a) evidence that automated collaboration assessments fall within the range of expert variability; (b) an architecture in which LLM-generated representations serve as both user-facing artifacts and indexed retrieval infrastructure for an AI agent; (c) an empirical evaluation showing that artifact-grounded responses are rated significantly higher than transcript-only responses on groundedness, analytical depth, and helpfulness.

%% file: 2_related_work.tex
\section{Related Work}
\subsection{Collaboration Literacy and Assessment}
Collaboration literacy includes the skills required to work effectively in group settings, including adapting to evolving collaborative dynamics~\cite{worsley2020towards}. Unlike domain knowledge assessed through standardized instruments, collaboration literacy manifests through contextually embedded and temporally distributed interactive behaviors~\cite{graesser2017complex,wise2011analyzing}, making collaboration quality resistant to point-in-time measurement and difficult to assess at scale~\cite{lamsa2021we}. 

Prior approaches to assessing collaboration have relied on behavioral indicators such as verbal participation rates and gesture/gaze patterns~\cite{cukurova2020modelling,schneider2016real}. While these behavioral metrics capture important dimensions of collaboration, the semantic content of discourse---how ideas are proposed, built upon, and integrated---requires a distinct form of interpretive analysis. An earlier line of work has developed rating schemes for assessing collaboration quality in CSCL settings, identifying dimensions such as mutual understanding, information pooling, and task orientation~\cite{meier2007rating,kahrimanis2009assessing}. However, these frameworks are typically designed for structured, technology-mediated problem-solving tasks and often require access to extensive process traces. In contrast, the 7C  dimensions from Worsley et al.~\cite{worsley2021designing} were derived from the construct of collaboration literacy rather than task-specific process behaviors, making them applicable to assessing collaboration quality across diverse and flexible discussion contexts. Yet applying the 7C framework, like other qualitative assessment approaches, still relies on labor-intensive manual analysis that limits scalability~\cite{feng2025analyzing,parfenova2025text}, motivating our use of LLMs to explore automated collaboration assessment.

\subsection{LLMs for Educational Assessment}
LLMs have shown promise across educational assessment tasks, including essay scoring~\cite{pack2024large}, feedback generation~\cite{dai2024assessing}, and knowledge tracing~\cite{scarlatos2025exploring}. Studies comparing LLM-generated assessments to human raters found that LLMs demonstrated stronger convergent validity in multi-dimensional contexts~\cite{wang2025evaluating}, though performance varied with task types and analytic rubrics. While most work focuses on individual student outputs where rubric criteria map to identifiable textual features, collaboration assessment poses distinct challenges: it targets interaction processes rather than individual outputs, and requires inferring group dynamics from dialogue where collaborative behaviors are often implicit~\cite{de2006content,strijbos2006content}. This makes collaboration assessment both a harder test case for LLMs and a domain where automated support could be particularly valuable given the difficulty of manual analysis at scale. Our work evaluates whether LLMs can produce reliable collaboration assessments using a multi-dimensional framework where even human experts show variable agreement.

\subsection{Retrieval-Augmented Generation for Educational Applications}
RAG architectures ground LLM responses in retrieved documents, reducing hallucination and improving factual accuracy~\cite{lewis2020retrieval}. Educational applications have included tutoring systems that retrieve relevant course materials and educational Q\&A systems that surface lecture content~\cite{li2025retrieval}. While most systems retrieve from human-authored corpora, CLARA instantiates a pipeline where one LLM pass produces structured knowledge that supports a second pass of agentic reasoning.

%% file: 3_system_design.tex
\section{System Design}
\begin{figure}[ht]
    \centering
    \includegraphics[width=\linewidth]{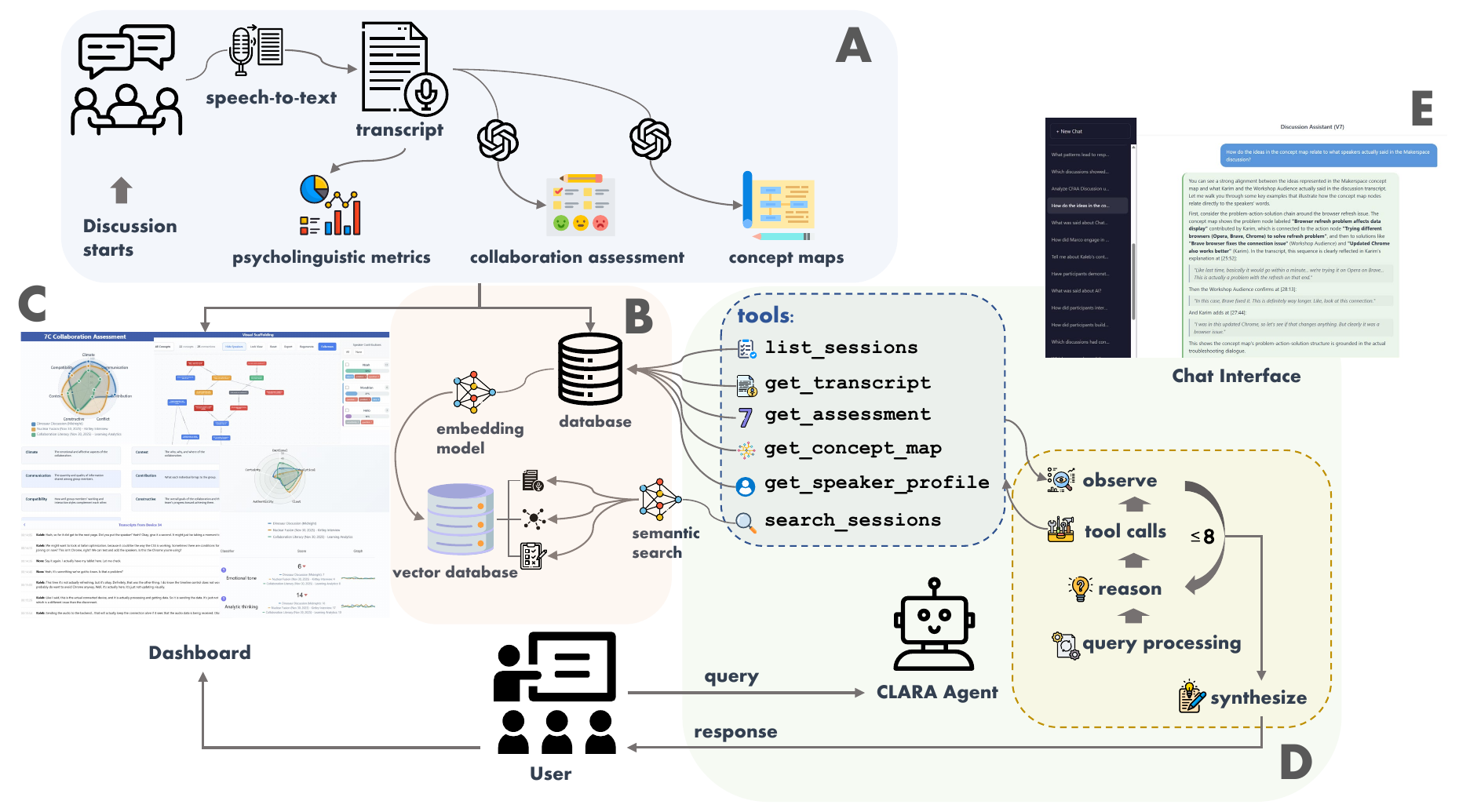}
    \caption{\textbf{System Workflow.} \textbf{(A)} CLARA transcribes discussion in real time, calculates psycholinguistic metrics, and prompts the LLM to produce collaboration assessments and concept maps; \textbf{(B)} Transcripts and artifacts are embedded and indexed in distinct database collections; \textbf{(C)} Users can interact with LLM-produced artifacts through the dashboard; \textbf{(D)} The CLARA Agent reasons and iteratively calls tools to query different artifacts for synthesizing responses; \textbf{(E)} The interface where users chat with the agent.}
    \label{fig:workflow}
\end{figure}

CLARA supports collaboration literacy by turning discussions into a set of artifacts that learners and educators can review and that an AI agent can search, as illustrated in Figure \ref{fig:workflow}. A central motivation for this design is to support users and the agent in reasoning over shared representations through conversational interaction. In addition, educators often query using terms that are not necessarily spoken verbatim in student dialogue, such as ``communication quality'', ``systems thinking'', and ``productive conflict''. When a query intrinsically spans multiple discussions, e.g., ``Which groups demonstrated effective mutual learning?'', the system must semantically search over all discussions and identify which are the most relevant. By indexing artifacts that explicitly summarize higher-level constructs, CLARA bridges this vocabulary gap and increases the chance that a conceptual query retrieves the right discussion.

\subsection{Semantic Artifact Generation and Indexing}
Audio streams sent to the server are transcribed independently per device using Google's Speech-to-Text model Chirp 2\footnote{\url{https://docs.cloud.google.com/speech-to-text/docs/models/chirp-2}}, which supports word-level timestamps and produces speaker-attributed text. Psycholinguistic metrics, such as \textit{analytic thinking} and \textit{certainty}, are computed at utterance level using dictionary-based methods and visualized in the dashboard, following Worsley et al.~\cite{worsley2021designing}. After a session ends, CLARA generates two semantic artifacts from the finalized transcript with OpenAI's GPT-4.1\footnote{\url{https://platform.openai.com/docs/models/gpt-4.1}}; both are shown in the dashboard and stored in machine-queryable form. 

\paragraph{Concept map.}The concept map represents the discussion as a graph capturing concepts introduced by the group and relationships between them. We defined these categories in consultation with two researchers in collaborative learning analytics to capture the epistemic structure of common collaboration activities. Concept types include \textit{idea}, \textit{question}, \textit{hypothesis}, \textit{example}, \textit{problem}, \textit{solution}, \textit{goal}, \textit{uncertainty}, \textit{conclusion}, and \textit{action}; Relationship types include \textit{supports}, \textit{builds on}, \textit{challenges}, \textit{exemplifies}, \textit{answers}, \textit{similar to}, \textit{leads to}, \textit{contrasts}, and \textit{relates to}. Guided by these categories and a structured output schema, the LLM generates concept maps zero-shot from the transcript. Using fixed types and relations keeps maps consistent across sessions and makes it easier to interpret and query them. Internally, the map is stored as a text-serializable graph (nodes, edges, identifiers), which the dashboard renders as an interactive visualization. Users can explore the map by zooming in and out, and opening linked transcript to see where a concept came from.

\paragraph{Collaboration assessment.}
CLARA produces a collaboration assessment across seven dimensions (the 7C framework) derived from Worsley et al.~\cite{worsley2021designing}. For each dimension, the assessment includes \textbf{(a)} a numeric score; \textbf{(b)} brief analysis describing observed patterns to explain the score; and \textbf{(c)} key evidence excerpts. In the dashboard, users can read the analysis and then jump into the transcript around the evidence to confirm, contextualize, or challenge the interpretation. 

The definitions of the seven dimensions were provided to the LLM to guide its generation of collaboration assessments: \textit{Climate:} the emotional and affective aspects of the collaboration; \textit{Communication:} the quantity and quality of information shared among group members; \textit{Compatibility:} how well group members' working and interaction styles complement each other; \textit{Conflict:} students' approaches to handling challenging or contentious situations that arise; \textit{Context:} the who, why, and where of the collaboration; \textit{Contribution:} what each individual brings to the group; \textit{Constructive:} the overall goals of the collaboration and the team's progress toward achieving them. 

\paragraph{Artifact indexing.}
Once transcripts or artifacts are created or updated, CLARA indexes them into ChromaDB\footnote{\url{https://www.trychroma.com/}} using OpenAI's \texttt{text-embedding-3-large}\footnote{\url{https://platform.openai.com/docs/models/text-embedding-3-large}}. Each artifact is indexed as a single document within its respective collection. No chunking is required because artifacts fall comfortably within the embedding model's 8,192-token context window. Separating collections by artifact type makes retrieval more controllable and supports mixed-initiative use: the system can search within a specific representation when the question calls for it, and users can restrict which representations the agent is allowed to consult.

\subsection{Agentic Workflow}
We built a ReAct agent~\cite{yao2022react} that gathers evidence through a structured loop of up to eight iterations, using LangGraph\footnote{\url{https://www.langchain.com/langgraph}} with GPT-4.1-mini\footnote{\url{https://platform.openai.com/docs/models/gpt-4.1-mini}}. At each step, the agent reviews the evidence it has accumulated, reasons about what additional information is needed to address the query, and selects tools to surface relevant information across the database, with six tools available: 
\begin{itemize}
\item\texttt{search\_sessions} performs semantic search across all indexed artifact collections, returning the most relevant discussions;
\item\texttt{list\_sessions} returns available discussion sessions and their metadata;
\item\texttt{get\_transcript} fetches discussion transcripts with psycholinguistic metrics; 
\item\texttt{get\_concept\_map} fetches discussion concept maps;
\item\texttt{get\_assessment} fetches scores, analyses, and evidence of a discussion's collaboration assessment;
\item\texttt{get\_speaker\_profile} retrieves information about an individual participant's contributions and interaction patterns across sessions, including participation share, concept contributions, and psycholinguistic metrics.
\end{itemize}

The agent iterates until it has enough evidence to produce a response specific and traceable to stored artifacts. Once retrieval is complete, a separate synthesis call generates the final response. Not all queries require calling the \texttt{search\_sessions} tool for semantic search---when users ask about a specific discussion, the agent can directly fetch its artifacts. The architecture supports the full range from targeted retrieval to open-ended analytical exploration.

\paragraph{Mixed-initiative controls.}
Users can constrain retrieval by choosing which artifact types to include or exclude (e.g., transcript + concept map only). This supports comparison across representations and gives users a way to set aside an artifact when they consider it unhelpful for a particular question.

%% file: 4_evaluation.tex
\section{Evaluation}
We evaluate CLARA along three dimensions. First, we examine whether LLM-generated 7C collaboration assessments align with assessments produced independently by human experts. Second, we compare agent responses generated with full artifact access against a transcript-only baseline to test whether artifacts improve response quality. Third, we evaluate retrieval performance under four conditions with different configurations of access to the artifacts.

\subsection{Study Context and Procedure}
\paragraph{Data collection.}
Approved by the university's IRB, we collect data from a diverse set of collaborative learning settings, from workshops with child participants to classroom discussions with college students, spanning technical debugging, policy analysis, generative AI, and exploratory science conversations. A \textit{session} refers to a single meeting event; participants work in parallel small groups, and each group's conversation constitutes a \textit{discussion} within each session. Two participants attended multiple sessions. The full dataset comprises 12 sessions containing 35 discussions. Session lengths ranged from 11.9 to 137.3 minutes, while individual processed discussions were generally under 25 minutes.

\paragraph{Raters and procedure.}
Four education researchers served as human experts for analyzing discussions with the 7C assessment framework. One was excluded after calibration review, leaving each discussion analyzed by 2 or 3 of the remaining experts. They received rubrics with dimension definitions, and rated independently, following the same format used to prompt the LLM. No rater saw the LLM's output; the LLM generated its own assessment from the same transcript.

For rating agent responses, we invited four external raters with backgrounds in education, including a program coordinator at a continuing education college, a high school teacher, an instructional designer, and a graduate student researcher in learning analytics who participated in a session as a facilitator. We walked the external raters through the system and showed them all available artifacts. Critically, they were blind to the two-condition design. All raters rated each of 46 agent responses independently across five dimensions.

\subsection{Evaluating Automated Collaboration Assessment}
\paragraph{Method.} We used 10 discussions for human experts analyzing collaboration. For each of the 10 discussions, human raters wrote complete 7C assessments independently, scoring each dimension on a 0--100 scale with a written analysis and key evidence. We compare human and LLM assessments at the score level and at the level of the written analyses.

Since the rating design produces missing cells, we use Krippendorff's alpha (interval level, 10,000 bootstrap iterations)~\cite{krippendorff2018content} rather than ICC to measure inter-rater agreement. We compute alpha among human raters per dimension, then recompute with the LLM included as an additional analyst. We also report Spearman's rank correlation coefficient (Spearman's $\rho$) between human mean scores and LLM scores across the 10 discussions, and the mean absolute difference (MAD) on the 100-point scale for human-human and human-LLM pairs. All measures are reported per dimension and aggregated across all 70 dimension-level assessments (10 discussions $\times$ 7 dimensions).

To assess whether the LLM attends to similar phenomena as human raters, we compare the written analyses and key evidence fields using an LLM-based structured evaluation~\cite{zheng2023judging}. For each dimension of each discussion, an independent judge model (Anthropic's Claude Sonnet 4, \texttt{claude-sonnet-4-20250514}\footnote{\url{https://www.anthropic.com/claude/sonnet}}) receives a pair of assessments, including both the analysis and key evidence fields, and rates two criteria on a 1--5 scale: \textbf{(1) behavioral alignment} (whether both analyses identify the same collaborative behaviors and reach similar conclusions), and \textbf{(2) evidence correspondence} (whether both analyses draw on similar moments from the discussion). A blank key evidence field, indicating insufficient evidence of the dimension, was treated as agreement if both analysts reached a similar conclusion. We run the comparison on all human-human pairs and all human-LLM pairs. Each comparison was run three times with analyst presentation order randomized; scores were averaged across runs. As a supplementary measure, we also compute pairwise cosine \textbf{embedding similarity} between analysis texts using OpenAI's \texttt{text-embedding-3-large} for all human-human and human-LLM pairs per dimension. We apply Mann-Whitney $U$ tests to compare the two distributions.

\begin{table}[t]
\centering
\small
\caption{Score-level agreement between human raters (H-H) and between humans and the LLM (H-LLM) across seven dimensions.}
\label{tab:7c_scores}
\begin{tabular}{l cc cc cc c}
\toprule
 & \multicolumn{2}{c}{Mean (SD)} & \multicolumn{2}{c}{Krippendorff's $\alpha$} & \multicolumn{2}{c}{MAD} & \\
\cmidrule(lr){2-3} \cmidrule(lr){4-5} \cmidrule(lr){6-7}
Dimension & Human & LLM & H-H & H-LLM & H-H & H-LLM & Spearman's $\rho$ \\
\midrule
Climate       & 76.4 (8.1)  & 78.5 (7.8)  & .300 & .412 & 7.89  & 6.88  & .711* \\
Communication & 72.3 (7.6)  & 78.3 (6.3)  & .525 & .462 & 6.78  & 6.79  & .760* \\
Compatibility & 76.8 (11.4) & 69.5 (5.5)  & $-.085$ & .048 & 12.56 & 8.46 & .585 \\
Contribution  & 69.3 (9.8)  & 67.0 (9.2)  & .567 & .472 & 7.22  & 9.04  & .475 \\
Constructive  & 71.9 (8.8)  & 69.0 (10.8) & .465 & .513 & 6.39  & 7.29  & .654* \\
Context       & 56.5 (17.2) & 63.0 (10.1) & .112 & .159 & 20.00 & 13.54 & .259 \\
Conflict      & 47.0 (19.4) & 53.0 (12.5) & .731 & .673 & 10.89 & 9.50  & .765** \\
\midrule
Overall       &  &  & .637 & .639 & 10.25 & 8.79  & .701*** \\
\bottomrule
\multicolumn{8}{l}{\footnotesize $^{*}p<.05$, $^{**}p<.01$, $^{***}p<.001$. Overall $\rho$: across all 70 dimension-level assessments.}
\end{tabular}
\end{table}

\begin{table}[t]
\centering
\small
\caption{Analytical alignment between human-human and human-LLM analysis pairs, assessed by an independent LLM judge model on two criteria (1--5 scale) and by embedding similarity. No \textit{overall} differences were significant on the two judge-rated criteria.}
\label{tab:7c_alignment}
\begin{tabular}{l cc cc cc}
\toprule
 & \multicolumn{2}{c}{Behavioral Align.} & \multicolumn{2}{c}{Evidence Correspond.} & \multicolumn{2}{c}{Embedding Sim.} \\
\cmidrule(lr){2-3} \cmidrule(lr){4-5} \cmidrule(lr){6-7}
Dimension & H-H & H-LLM & H-H & H-LLM & H-H & H-LLM \\
\midrule
Climate       & 3.40 & 3.55 & 2.77 & 2.62 & .478 & .530 \\
Communication & 3.63 & 3.00* & 2.87 & 2.23 & .440 & .478 \\
Compatibility & 3.23 & 3.40 & 2.80 & 2.60 & .411 & .454 \\
Contribution  & 3.77 & 3.90 & 2.77 & 2.80 & .461 & .539 \\
Constructive  & 3.67 & 3.03* & 2.90 & 2.37 & .438 & .405 \\
Context       & 2.67 & 2.77 & 2.40 & 2.40 & .388 & .434 \\
Conflict      & 4.17 & 4.37 & 3.27 & 3.53 & .477 & .510 \\
\midrule
Overall       & 3.50 & 3.43 & 2.82 & 2.65 & .442 & .479 \\
\bottomrule
\multicolumn{7}{l}{\footnotesize $^{*}p<.05$ (Mann-Whitney $U$ test, per-dimension comparison).}
\end{tabular}
\end{table}

\paragraph{Results of score-level agreement.} Table~\ref{tab:7c_scores} reports agreement between human raters and the LLM across all collaboration dimensions. Human inter-rater reliability varied by dimension: Conflict showed the strongest agreement ($\alpha = .731$), followed by Contribution (.567) and Communication (.525); Compatibility ($-.085$) and Context (.112) showed near-zero agreement, indicating that these dimensions were difficult even for trained human experts to assess consistently. Overall human agreement was moderate ($\alpha = .637$, 95\% CI $[.452, .760]$).

\textbf{LLM does not reduce agreement when added as an additional rater.} Overall $\alpha$ was essentially unchanged ($.637$ vs.\ $.639$), and on dimensions where human agreement was weak, including Climate, Compatibility, Context, $\alpha$ increased when the LLM was included ($.300$ to $.412$, $-.085$ to $.048$, and $.112$ to $.159$, respectively). The LLM's scores fell within the range of human variability.

\textbf{LLM-produced scores' rank-ordering of discussions tracked human consensus on most dimensions} (overall $\rho= .701$, $p < .001$). Conflict ($\rho= .765$), Communication ($\rho= .760$), and Climate ($\rho= .711$) showed the strongest rank-order agreement. Context was the weakest ($\rho= .259, p = .470$), consistent with the low human agreement on that dimension. Compatibility ($\rho= .585, p = .076$) and Contribution ($\rho= .475, p = .165$) fell in between.

\textbf{In absolute terms, LLM-produced scores deviated less from the human mean than individual human scores deviated from each other} (overall MAD: $8.79$ vs.\ $10.25$ on the 100-point scale). This held on five of seven dimensions; Contribution ($9.04$ vs.\ $7.22$) and Constructive ($7.29$ vs.\ $6.39$) were exceptions where human-LLM deviation slightly exceeded human-human deviation. The largest MAD differences appeared on dimensions with high human disagreement: Context ($13.54$ vs.\ $20.00$) and Compatibility ($8.46$ vs.\ $12.56$). The LLM's score distributions were narrower than human distributions on most dimensions (e.g., Conflict SD: $12.5$ vs.\ $19.4$; Compatibility SD: $5.5$ vs.\ $11.4$), reflecting a tendency toward moderate scores rather than extreme judgments.

\paragraph{Results of analytical alignment.} \textbf{No significant difference between human-human and human-LLM pairs was found on either criterion}  (Table~\ref{tab:7c_alignment}; behavioral alignment: $3.50$ vs.\ $3.43$, $p = .370$; evidence correspondence: $2.82$ vs.\ $2.65$, $p = .277$). Two dimensions showed significantly lower human-LLM behavioral alignment: Communication ($3.00$ vs.\ $3.63$, $p = .042$) and Constructive ($3.03$ vs.\ $3.67$, $p = .020$), indicating that on these dimensions the LLM sometimes characterized collaborative behaviors differently from experts despite assigning comparable scores. Evidence correspondence did not differ significantly on any individual dimension.

Semantic similarity between texts was comparable for human-human and human-LLM pairs (Table~\ref{tab:7c_alignment}; cosine similarity: $.442$ vs.\ $.479$, $p = .0006, r = .193$). The small but statistically significant difference favored human-LLM pairs, indicating that the LLM's analyses were at least as semantically aligned with individual human analyses as experts were with each other.

\subsection{Evaluating Agent Responses}
\paragraph{Method.} We consulted learning analytics experts and drew 21 queries from a query pool of 56 queries representing analytical questions educators often pose: questions about discussion content, collaboration quality, participant contributions, and cross-session patterns. 

To isolate the value of artifact-based retrieval and reasoning for the \textbf{full} agent, we implement a \textbf{baseline} agent with the same architecture and prompting but restricted access to transcripts only. Its \texttt{search\_sessions} tool searches only the transcript collection and it cannot fetch concept maps or collaboration assessments. In our evaluation, 21 queries were tested on both agents, yielding 42 responses. Four queries referenced specific artifact types and were thus tested only on the full agent to keep the comparison fair, yielding 46 responses in total. Raters evaluated all responses blind to condition on 5-point Likert scales across five dimensions: \textit{Accuracy} (factual correctness regarding mentioned discussions), \textit{Relevance} (how directly the response addresses the query), \textit{Groundedness} (whether claims cite specific evidence rather than being generic), \textit{Analytical Depth} (insight beyond surface description), and \textit{Helpfulness} (practical value to a learner or educator in terms of collaboration literacy).

For the 21 pairs, we compare mean rater scores per response using Wilcoxon signed-rank tests, reporting effect sizes as matched-pairs rank-biserial correlation ($r$). We also report an overall quality score (mean across dimensions). For the 4 artifact-specific queries, we report descriptive statistics only.
\begin{table}[t]
\centering
\small
\caption{Agent response quality ratings (1--5 Likert scale) by four expert raters across 21 matched query pairs. ICC(2,1) computed on all 42 paired responses. Effect size $r$ is matched-pairs rank-biserial correlation from Wilcoxon signed-rank tests.}
\label{tab:agent_responses}
\begin{tabular}{l c cc cc}
\toprule
 & & \multicolumn{2}{c}{Mean Rating} & & \\
\cmidrule(lr){3-4}
Dimension & ICC(2,1) & Full & Baseline & $r$ & $p$ \\
\midrule
Accuracy         & .590 & 4.24 & 3.93 & .476 & .054 \\
Relevance        & .598 & 4.32 & 3.79 & .528 & .033* \\
Groundedness     & .673 & 4.46 & 3.74 & .693 & .005** \\
Analytical Depth & .551 & 4.17 & 3.42 & .723 & .004** \\
Helpfulness      & .441 & 4.02 & 3.48 & .737 & .006** \\
\midrule
Overall          & .690 & 4.24 & 3.67 & .792 & .001** \\
\bottomrule
\multicolumn{6}{l}{\footnotesize $^{*}p<.05$, $^{**}p<.01$.}
\end{tabular}
\end{table}

\paragraph{Results.} Inter-rater reliability across the 42 paired responses was moderate to good (Table~\ref{tab:agent_responses}). Groundedness showed the highest single-rater agreement (ICC(2,1) = .673), followed by Relevance (.598) and Accuracy (.590). Analytical Depth (.551) and Helpfulness (.441) were lower, consistent with these dimensions requiring more subjective judgment. Reliability of the averaged ratings used in subsequent analyses was acceptable across all dimensions (ICC(2,k) $\geq .759$).

\textbf{The full agent received significantly higher ratings than the baseline on four out of five dimensions and overall} (Table~\ref{tab:agent_responses}). The largest effects appeared on Helpfulness ($r = .737$, $p = .006$), Analytical Depth ($r = .723$, $p = .004$), and Groundedness ($r = .693$, $p = .005$), dimensions where access to structured analytical representations is expected to matter most. Relevance was also significantly higher for the full agent ($r = .528$, $p = .033$). The overall comparison was significant ($r = .792$, $p = .001$), with the full agent averaging 4.24 and the baseline 3.67 on the 5-point scale. In addition, the four artifact-specific queries tested only on the full agent were rated consistently high (M = 4.00--4.25 across dimensions).

Accuracy did not differ significantly between conditions ($r = .476$, $p = .054$). Both agents drew on the same transcripts and produced factually comparable responses; the advantage of artifact access appeared not in factual correctness but in the depth and grounding of the analysis, requiring \textit{contextually} characterizing \textit{how} the group collaborated. While the baseline agent could reason about collaboration from raw dialogue, the full agent leveraged artifacts as additional analytical lenses, scaffolding its reasoning with pre-structured representations to better identify epistemic relationships and contextualize collaboration quality.  

\subsection{Evaluating Artifact-Grounded Retrieval}
\paragraph{Method.} The agent's \texttt{search\_sessions} tool performs semantic search across all indexed collections using reciprocal rank fusion (RRF), which merges ranked lists into a unified ranking, prioritizing discussions that rank highly across multiple artifact types. To inspect the contribution of different artifacts to retrieval quality, we evaluated four configurations: transcript-only, transcript with concept maps, transcript with 7C collaboration assessments, and all artifacts.

We curate 30 evaluation queries from the query pool: 10 \textbf{direct} queries with terms appearing in transcript text (e.g., ``What was said about AI?'') and 20 \textbf{analytical} queries with evaluative or interpretive vocabulary that does not typically appear in speakers' speech (e.g., ``Which discussions showed strong communication quality?''). For each query, we identified ground truth discussions that address the query, through manual review by two members of the research team. Queries were curated to ensure that multiple sessions were relevant per query, allowing for graded recall measurement. For each query-configuration pair, we ran the retrieval module and recorded the ranked list of returned discussions. We report \textsc{Recall@5} (proportion of relevant sessions appearing in the top 5 results), \textsc{Recall@10}, and Mean Reciprocal Rank (\textsc{MRR@5}, the reciprocal of the rank of the first relevant result within the top 5). Metrics are averaged across queries within each category.
\begin{table}[t]
\centering
\small
\caption{Retrieval performance across artifact configurations and query types.}
\label{tab:retrieval}
\begin{tabular}{l ccc ccc}
\toprule
 & \multicolumn{3}{c}{\textsc{Direct}} & \multicolumn{3}{c}{\textsc{Analytical}} \\
\cmidrule(lr){2-4} \cmidrule(lr){5-7}
Condition & Recall@5 & Recall@10 & MRR@5 & Recall@5 &  Recall@10 &  MRR@5 \\
\midrule
Transcript   & .867 & .967 & 1.000 & .371  & .615 & .727 \\
Transcript + CM   & .900 & .933 & 1.000 & .534          & .739 & .706 \\
Transcript + 7C & .933 & .967 & 1.000 & .563          & .774 & .817 \\
All Artifacts     & .900 & .967 & 1.000 & \textbf{.739} & \textbf{.804} & \textbf{.942} \\
\bottomrule
\end{tabular}
\end{table}

\paragraph{Results.} Table~\ref{tab:retrieval} reports retrieval performance under four artifact configurations, separated by query type. \textbf{On direct queries, all configurations performed comparably} (Recall@5: .867--.933; Recall@10: .933--.967; MRR@5: 1.000 across conditions). When query terms appear in transcripts, the retrieval layer finds relevant sessions regardless of which collections are available.

\textbf{Performance diverged on analytical queries.} Transcript-only retrieval found fewer than half of the relevant sessions (Recall@5 $= .371$), reflecting the mismatch between evaluative terms in the queries (e.g., ``communication quality'', ``systems thinking'') and the informal language of student speech. Adding concept maps improved coverage (Recall@5 = .534), as did adding 7C collaboration assessments (Recall@5 $= .563$). Combining all artifact types yielded the strongest performance (Recall@5 $= .739$, Recall@10 $= .804$, MRR@5 $= .942$).

The 7C assessments contributed slightly more than concept maps did ($.563$ vs. $.534$). This likely reflects the nature of each representation: 7C collaboration assessments are written in evaluative language that overlaps with how educators phrase analytical questions, while concept maps encode content structure like concepts, relationships, and their types, which, though useful, are less closely aligned with evaluative vocabulary. The two were complementary; combining them yielded higher recall than either alone (Recall@5 $= .739$ vs. $.563$ and $.534$).

%% file: 5_discussion.tex
\section{Discussion}
\subsection{Automated Assessment for Collaboration Literacy}
The evaluation results of 7C collaboration assessment suggest that LLM-generated collaboration assessments can reach a level of alignment with expert judgment that is practically useful, even if imperfect. On most dimensions, LLM scores fell within the spread of human rater scores, and the structured comparison of written analyses indicated that the LLM identified similar collaborative behaviors and drew on similar evidence. These findings are consistent with recent work demonstrating that LLMs can perform comparably to or better than human raters on complex, multi-dimensional educational assessment tasks~\cite{wang2025evaluating,dai2024assessing}. This does not mean the LLM is interchangeable with a trained analyst; there were dimensions where agreement was weaker, and the qualitative analyses occasionally differed in emphasis or specificity. As a structured entry point for interpreting a discussion, however, the assessments appear worth engaging with.

Beyond measurement, the assessments can act as a catalyst for building collaboration literacy. One barrier to developing collaboration skills is that the feedback loop is slow: structured reflection on how the group collaborated typically requires a time-consuming process of observation, analysis, and debriefing. The automated assessment streamlines this process. We believe that these assessments are better understood as scaffolds than as ground truth---their value lies not in being definitively correct but in giving educators and learners a structured characterization to react to, agree with, push back on, or use as a starting point for deeper reflection. The artifacts make collaboration visible and discussable in ways that raw transcripts or simple participation metrics do not. 

\subsection{LLM Artifacts for Semantic Retrieval and Grounding}
A transcript contains everything that was said, so in principle, an agent with transcript access has all the information it needs. Nevertheless, the full agent produced responses that raters judged as more grounded and analytically deeper, and transcript-only retrieval struggled to surface relevant discussions for analytical queries---those framed in evaluative language that does not appear verbatim in discussions. This confirms the effectiveness of utilizing structured knowledge produced from the transcript for agent retrieval and synthesis.

Raw transcripts are linear and often noisy with false starts, digressions, and overlapping ideas. Extracting analytical insight from them requires simultaneously identifying \textit{what matters} in the discourse, distilling it into analytical structures, and reasoning over the result. The extracted artifacts effectively offload the first two steps. A concept map would have already identified the key concepts and their relationships; a collaboration assessment would have already characterized collaboration quality with grounded analysis. The agent can then focus on synthesis and interpretation rather than re-deriving structure from raw text.

In some sense, the system is creating its own retrieval infrastructure: the artifacts generated in the first pass become the indexed knowledge that supports reasoning in the second pass. Whether more open-ended or less formalized representations would provide the same retrieval advantage is an open question.

\subsection{Educational Implications of Human-AI Sensemaking}
CLARA was designed around the idea that learners, educators, and AI should reason over the same representations. In practice, this means that when a learner or educator looks at a concept map and then asks the agent a question, both are working from a shared reference. The educator can point to a node and ask why it matters; the agent can cite a relationship from the same map in its response. This shared footing is what we mean by common ground---not in the full sense of mutual knowledge verified through grounding acts~\cite{clark1991grounding}, but in the more practical sense that both parties can refer to the same analytical objects.

Conventional analytics systems follow a generate-and-consume pattern: the AI generates outputs and the user consumes them; the user might trust or distrust the outputs, but there is no shared object to negotiate over. In CLARA, the artifacts sit between the user and the AI, which are visible to both, queryable by both, and interpretable by both. This suggests a design direction for AI-augmented analytics: rather than building systems where AI generates insights for users to accept or reject, we can build systems where AI generates representations that become shared working material. The representations are thus not endpoints but infrastructure that support ongoing inquiry from both sides.

\subsection{Limitations}
While the agent response and retrieval evaluations draw on the full dataset of 35 discussions, the collaboration assessment comparison relies on 10 discussions analyzed by human experts, a sample sufficient to identify meaningful patterns but not to make strong generalizations. For the agent comparison, ICC estimates should be interpreted with caution given the number of raters, and the paired comparison has relatively limited power to detect small effects. We report effect sizes throughout to ensure the results are interpretable despite these constraints. Extending the evaluation to a larger scale would be a natural next step.

%% file: 6_conclusion.tex
\section{Conclusion}
We presented CLARA, a system in which LLM-generated analytical artifacts,  including concept maps and collaboration quality assessments, serve a dual role: as visualized scaffolds for users to interact with and as indexed knowledge for an AI agent to reason over. Our evaluation showed that automated collaboration assessments fall within the range of expert variability, that artifacts nearly double retrieval recall on analytical queries by bridging student discourse and educator vocabulary, and that artifact-grounded agent responses are rated significantly higher on groundedness, analytical depth, and helpfulness.

Our findings point to a broader design pattern of generating structured representations that serve both human interpretation and AI reasoning, which may generalize beyond learning analytics to any setting where AI-produced analyses are both presented to users and utilized to support downstream human-AI sensemaking. We also note that the more consequential question is whether these artifacts and agent responses change how learners and educators reflect on and respond to their collaboration practice, a question that requires large-scale classroom deployment studies. 